\documentclass{article}

\usepackage{fullpage}

\usepackage{hyperref}
\usepackage{amsmath}
\usepackage{amssymb}
\usepackage[vlined]{algorithm2e}
\usepackage{graphics}
\usepackage{graphicx}
\usepackage{epstopdf}
\usepackage{subfigure}
\usepackage{pifont}
\usepackage{multicol}
\usepackage{tikz}
\usepackage{pgfplots}

\def\bp{\noindent {\it Proof.}\ }
\def\ep{\hfill $\Box$} 

\def\be{\begin{equation}}
\def\ee{\end{equation}}

\newtheorem{prop}{Proposition}

\begin{document}

\title{Learning Graph Representations by Dendrograms}

\author{Thomas Bonald, Bertrand Charpentier}
\date{\today}

\maketitle

\begin{abstract}
Hierarchical graph clustering is a common technique to reveal the 
multi-scale structure of  complex networks.  
We propose a novel  metric for assessing the quality of a hierarchical clustering. This metric reflects the ability to reconstruct the graph from the dendrogram, which encodes the hierarchy. The optimal representation of the graph defines a class of reducible  
 linkages leading to regular dendrograms by greedy agglomerative clustering.
\end{abstract}

\section{Introduction}

Many datasets have a  graph structure. Examples include infrastructure networks, communication networks, social networks, databases and co-occurence networks, to quote a few. These graphs often exhibit a complex, multi-scale  structure where each node belong to many groups of nodes, so-called clusters, of different sizes. 

Hierarchical graph clustering is a common technique to reveal the 
multi-scale structure of  complex networks.  
Instead of looking for a single partition of the set of nodes, as in usual clustering techniques, the graph is represented by a hierarchical structure known as a dendrogram, which can then be used to find relevant clusterings at different resolutions, by suitable cuts of this dendrogram.

We address  the issue of the  
 {\it quality} of a dendrogram representing a graph. For usual clustering, the quality of a partition is  commonly assessed through its modularity, which corresponds to the proportion of edges within clusters, compared to that of a random graph  \cite{newman2004}. 
 For hierarchical clustering, 
 a cost function has recently been proposed by Dasgupta \cite{dasgupta2016cost} and extended by Cohen-Addad and his co-authors \cite{cohen2018hierarchical}; it can be viewed as  the expected size of the smallest cluster (as induced by the hierarchy) containing two random nodes.
In this paper,  we propose a quality metric based on the ability to reconstruct the graph from the dendrogram. The optimal representation of the graph defines a class of reducible  
 linkages leading to regular dendrograms by greedy agglomerative clustering.

In the next section,  we introduce the sampling distributions of nodes and clusters induced by the graph; these play a central node in our approach.
We then formalize the problem of graph representation by a dendrogram. Our quality metric follows from the characterization of the optimal solution  in terms of   graph reconstruction. 
The corresponding hierarchical graph clustering algorithms  are then presented and interpreted in terms of modularity.

\section{Sampling distribution}

Consider a weighted, undirected, connected graph $G = (V,E)$ of $n$ nodes, without self-loops. Let $w(u,v)$ be equal to the weight of edge $u,v$, if any, and  to 0 otherwise.
We refer to the weight of node $u$ as: 
$$
w(u) = \sum_{v\in V} w(u,v).
$$
We denote by $w$ the total weight of nodes: 
$$
w = \sum_{u\in V}w(u) = \sum_{u,v\in V}w(u,v). 
$$

Similarly, for any  sets $A,B\subset V$, let
$$
w(A,B) = \sum_{u\in A, v\in B} w(u,v),
$$
and 
$$
w(A) = \sum_{u\in A} w(u).
$$

\paragraph{Node sampling.}
The  weights induce a probability distribution on node pairs:
$$
\forall u,v\in V,\quad  p(u,v) = \frac{w(u,v)}{w},
$$
with marginal distribution:
$$
\forall u\in V,\quad p(u) =\sum_{v\in V} p(u,v) =  \frac{w(u)}{w}.
$$
The joint distribution $p(u,v)$ is  the relative frequency of moves from node $u$ to node $v$  by a random walk in the graph, with transition probability:
$$
\forall u,v\in V,\quad   p(v|u) = \frac{p(u,v)}{p(u)} =  \frac{w(u,v)}{w(u)}.
$$

\paragraph{Cluster sampling.}
For any partition ${\cal P}$ of $V$ into clusters, the  weights induce a probability distribution on cluster pairs:
$$
\forall A,B \in {\cal P},\quad  p(A,B) = \frac{w(A,B)}{w},
$$
with marginal distribution:
$$
\forall A\in {\cal P},\quad p(A) =\sum_{B\in {\cal P}} p(A,B) =  \frac{w(A)}{w}.
$$
The joint distribution $p(A,B)$ is  the relative frequency of moves from cluster $A$ to cluster $B$  by the random walk.

\section{Representation by a dendrogram}

Assume the graph $G$ is represented by a  dendrogram, that is a rooted binary tree ${\cal T}$ whose  leaves are the nodes  of the graph, $V$. We denote by ${\cal I}$ the set of internal nodes of the tree ${\cal T}$. For each  $i\in {\cal I}$, a positive number $d(i)$ is assigned to node $i$, corresponding to its height  in the dendrogram. We assume that the dendrogram is regular in the sense that  $d(i) \ge d(j)$ if $i$ is an ancestor of $j$ in the tree.
For each $i\in {\cal I}$, there are two subtrees attached to node $i$, with sets of leaves $A$ and $B$; these sets uniquely identify the internal node $i$  so that we can write  $i = (A,B)$ and $d(i) = d(A,B)$.

\paragraph{Ultrametric.}
The dendrogram defines a metric on $V$: for each $u,v\in V$, we define  $d(u,v) = d(i)$  where $i$ is the closest common ancestor of $u$ and $v$ in the dendrogram. This is an ultrametric in the sense that:
$$
\forall u,v,x\in V,\quad d(u,v) \le \max(d(u,x), d(v,x)),
$$
Conversely, any ultrametric defines a dendrogram, which can be build from bottom to top by successive  peerings of the closest nodes. 

\paragraph{Graph representation.} 
A natural question is whether, given the dendrogram, it is possible to generate some graph $\hat G$ that is ``close" to the original graph $G$. Let $\pi$ be some probability distribution on $V$ representing the prior information known about the relative node weights: the distribution $\pi$ is uniform in the absence of  such information  and is equal to the sampling distribution $p$ for a perfect knowledge of the node weights. We look for the best representation of $G$ by a dendrogram $d$ in the sense of the reconstruction  of $G$ from $d,\pi$, which can be viewed as the  autoencoding scheme:
$$
G\longmapsto d,\pi \longmapsto \hat G.
$$

Since $d(u,v)$ can be interpreted as a distance between  $u$ and $v$, its inverse corresponds to a similarity. Thus we define the weight $\hat w (u,v)$ between any two nodes $u,v\in V$ in the graph $\hat G$ as:
$$
\hat w (u,v) =   \frac{\pi(u)\pi(v)}{d(u,v)} 1_{\{u\ne v\}}.
$$
Denoting by $\hat w$ the total weight, 
$$
\hat w = \sum_{u,v\in V} \hat w(u,v),
$$
we get the following node pair sampling distribution associated with  $\hat G$:
$$
\hat p(u,v) = \frac{\hat w(u,v)}{\hat w}.
$$
The distance between graphs $G$ and $\hat G$ can then be assessed through the Kullback-Leibler divergence between the respective sampling distributions,
$$
D(p||\hat p) = \sum_{u\ne v}p(u,v) \log\frac{ p(u,v)}{\hat p(u,v)}.
$$


\paragraph{Optimization problem.} Minimizing the  Kullback-Leibler divergence
$D(p||\hat p)$ in $\hat p$ is equivalent to minimizing the cost function:
$$
J(d) = \sum_{u\ne v}p(u,v)  \log d(u,v) + \log\left(\sum_{u\ne v} \frac {\pi(u)\pi(v)}{d(u,v)}\right),
$$
over all ultrametrics $d$ defined on $V$. 
 Observe that $J(\alpha d) = J(d)$ for any  $\alpha > 0$, so that the best ultrametric is defined up to a multiplicative constant.
Using the fact that $d(u,v) = d(i)$, where $i$ is the closest common ancestor of $u$ and $v$ in the dendrogram, we get:
\be\label{eq:cost2}
J(d) = \sum_{A,B: (A,B) \in {\cal I}}p(A,B) \log d(A,B) +  \log\left(\sum_{A,B: (A,B) \in {\cal I}} \frac {\pi(A)\pi(B)}{d(A,B)}\right),
\ee
with $$\pi(A) = \sum_{u\in A} \pi(u).$$
The problem of the best representation of $G$ by a dendrogram now reduces to the optimization problem:
\be\label{eq:opt}
\arg\min_d J(d),
\ee
over all ultrametrics $d$  on $V$.

\section{Optimal representation}

We seek to solve  the optimization problem \eqref{eq:opt}.

\paragraph{Optimal distances.}
We first assume that the underlying tree ${\cal T}$ of the ultrametric is given and look for the best corresponding distance $d$. Let ${\cal I}$ be the set of internal nodes of the tree. Then for each $(A,B)\in {\cal I}$, we get by the differentiation of \eqref{eq:cost2} in $d(A,B)$,
$$
\frac{p(A,B)}{d(A,B)} = \lambda \frac{\pi(A)\pi(B)}{d(A,B)^2},
$$
where
$$
\lambda = \left(\sum_{A,B:(A,B) \in {\cal I}} \frac {\pi(A)\pi(B)}{d(A,B)}\right)^{-1},
$$
that is 
\be \label{eq:linkage}
d(A,B) = \lambda \frac{\pi(A)\pi(B)}{p(A,B)}.
\ee

\paragraph{Optimal tree.} Replacing $d(A,B)$  by its optimal value \eqref{eq:linkage} in \eqref{eq:cost2},  
we deduce that the optimization problem \eqref{eq:opt} reduces to:
\be \label{eq:cost}
\arg\max_{\cal T}  \sum_{A,B: (A,B) \in {\cal I}}p(A,B) \log \frac{p(A,B)}{\pi(A) \pi(B)},
\ee
where ${\cal I}$ is the set of internal nodes of the tree ${\cal T}$. The dendrogram is then  fully
 determined by \eqref{eq:linkage}, for each internal node $(A,B) \in {\cal I}$.
The function to maximize in \eqref{eq:cost} is the Kullback-Leibler divergence between the cluster  pair distributions when nodes are  sampled from the edges and independently from the   distribution $\pi$, respectively. 
It provides a meaningful objective function for hierarchical clustering, that can be interpreted in terms of graph reconstruction.

A key difference between our 
 objective function \eqref{eq:cost} and the cost functions proposed in the literature \cite{dasgupta2016cost,cohen2018hierarchical} lies in the entropy term:
 $$
  \sum_{A,B: (A,B) \in {\cal I}}p(A,B) \log p(A,B).
 $$
Removing this  term yields the cost function:
$$
 \sum_{A,B: (A,B) \in {\cal I}}p(A,B) (\log\pi(A) + \log\pi(B)),
$$
to be compared with usual cost functions, of the form:  
$$
 \sum_{A,B: (A,B) \in {\cal I}}p(A,B) (\pi(A) + \pi(B)).
$$
When $\pi$ is the uniform distribution (no prior information on the node weights), these cost functions become respectively:
$$
\sum_{A,B: (A,B) \in {\cal I}}p(A,B) (\log |A| + \log |B|) \quad \text{and} \quad 
\sum_{A,B: (A,B) \in {\cal I}}p(A,B) (|A| + |B|).
$$
The latter is Dasgupta's cost function, equal to the expected size of the smallest cluster  containing two random nodes sampled from $p$. 

The optimization problem \eqref{eq:cost} is NP-hard, just like minimizing  Dasgupta's cost function is NP-hard \cite{dasgupta2016cost}. In the next section, we present  heuristics based on greedy agglomerative algorithms for finding approximate solutions to this optimization problem.

\section{Hierarchical clustering}
\label{sec:hierarchical}

The optimal distances   \eqref{eq:linkage} suggest a greedy algorithm for solving the optimization problem \eqref{eq:cost}. The algorithm consists in starting from $n$ clusters (one per node) and in successively merging the two closest clusters in terms of inter-cluster distance \eqref{eq:linkage}. This is a usual agglomerative algorithm with linkage (inter-cluster similarity):
\be \label{eq:linkage1}
\sigma(A,B) =  \frac{p(A,B)}{\pi(A)\pi(B)}.
\ee
The dendrogram is built from bottom to top, with distance $\sigma(A,B)^{-1}$ attached to the internal node $(A,B)$ resulting from the merge of  clusters $A,B$.
The agglomeration relies on the following update formula:

\begin{prop}
\label{prop:up}
We have:
$$
\sigma(A\cup B,C) =\frac{\pi(A)}{\pi(A\cup B)}{\sigma(A,C)} +\frac{\pi(B)}{\pi(A\cup B)}{\sigma(B,C)}.
$$
\end{prop}
\bp
We have:
\begin{align*}
\frac{\pi(A)}{\pi(A\cup B)} {\sigma(A,C)} +\frac{\pi(B)}{\pi(A\cup B)}{\sigma(B,C)} &= \frac 1 {\pi(A\cup B)\pi(C)}(p(A,C) + p(B,C)),\\
& = \frac {p(A\cup B,C)} {\pi(A\cup B)\pi(C)}= \sigma(A\cup B,C).
\end{align*}
\ep
\\

The update formula shows that  the linkage \eqref{eq:linkage1}  is reducible, in the sense that:
$$
\sigma(A\cup B,C)\le \max(\sigma(A,C),\sigma(B,C)).
$$
This inequality guarantees that the resulting  dendrogram is regular (the sequence of distances attached to successive internal nodes is non-decreasing) and that  the  corresponding distance on $V$  is an ultrametric. Moreover,  the search of the clusters to merge can be done through the nearest-neighbor chain to speed up the algorithm \cite{murtagh}.

\paragraph{Linkage.}
For $\pi$ the uniform distribution  (no prior information on the node weights), the linkage \eqref{eq:linkage1} is proportional to the usual average linkage:
$$
\sigma(A,B) \propto  \frac{w(A,B)}{|A||B|},
$$
corresponding to the density of the cut separating clusters $A$ and $B$.

For $\pi$ equal to $p$ (perfect knowledge  of the node weights), this is the linkage proposed in  \cite{paris}: 
$$
\sigma(A,B) =  \frac{p(A,B)}{p(A)p(B)},
$$
which can be interpreted in terms of sampling ratio as:
$$
\sigma(A,B) =  \frac{p(A|B)}{p(A)} =  \frac{p(B|A)}{p(B)}.
$$
We refer to this linkage as modular in view of its relationship with modularity.

\paragraph{Modularity.} The modularity of any partition ${\cal P}$ of the set of nodes $V$ is defined by \cite{newman2004}:
$$
Q =\sum_{C\in {\cal P}} \sum_{u,v \in C}    (p(u,v) - p(u)p(v)).
$$
This is the difference between the probabilities that two nodes belong to the same cluster when sampled from the edges and independently from the nodes, in proportion to their weights. The former sampling distribution depends on the graph while the latter  depends on the graph through the node weights only. 

A more general definition of modularity is:
$$
Q =\sum_{C\in {\cal P}} \sum_{u,v \in C}  (p(u,v) - \pi(u)\pi(v)).
$$
Now the node sampling distribution $\pi$ is any distribution with support $V$. For instance, it may be uniform (no prior information on the node weights) or equal to $p$ (the usual definition of modularity, where the information on the node weights is known)\footnote{Another common interpretation of modularity is the difference between the proportions of edge weights within clusters in the original graph and in some null model where nodes $u$ and $v$ are linked with probability $\pi(u)\pi(v)$;  for $\pi$ the uniform distribution  (no prior information on the node weights), the null model is an  Erd\H{o}s-R\'enyie graph while for $\pi = p$ (perfect knowledge of the node weights), the null model is the    configuration model.}.

Maximizing modularity usually provides a unique clustering. To explore the multi-scale structure of real graphs, it is common to introduce some positive resolution parameter $\gamma$ that controls the respective weights of  both terms in the definition of modularity \cite{reichardt2006,lambiotte2015,newman2016}. The modularity of partition ${\cal P}$ at resolution $\gamma$ is  defined by:
$$
Q_\gamma =\sum_{C\in {\cal P}} \sum_{u,v \in C} (p(u,v) -\gamma \pi(u)\pi(v)).
$$
When $\gamma \to 0$, the second term becomes negligible and the best partition ${\cal P}$ is trivial, with a single cluster equal to the set of nodes $V$.  When $\gamma \to +\infty$, the second term becomes preponderant and the best partition ${\cal P}$ has  $n$ clusters, one per node. Now the maximum resolution  beyond which the best partition ${\cal P}$ has  $n$ clusters is given by:
$$
\max_{u\ne v} \frac{p(u,v)}{\pi(u)\pi(v)}.
$$
 The first node pair to be merged (at maximum resolution) is that achieving this maximum, which is the closest pair in terms of linkage \eqref{eq:linkage1}. 

More generally,  given some clustering ${\cal C}$, the modularity of any coarser partition ${\cal P}$ at resolution $\gamma$ is:
$$
Q_\gamma = \sum_{C\in {\cal P}} \sum_{A,B\in {\cal C}:A,B \subset C} (p(A,B) -\gamma \pi(A)\pi(B)).
$$
Again, the best partition is ${\cal C}$ when $\gamma \to +\infty$ and the first cluster merge occurs at resolution:
$$
 \max_{A,B \in {\cal C}, A\ne B} \frac{p(A,B)}{\pi(A)\pi(B)}.
$$
The   cluster pair to be merged (at maximum resolution) is that achieving this maximum, which the closest pair $A,B\in{\cal C}$ in terms of linkage \eqref{eq:linkage1}. The agglomerative algorithm based on  linkage \eqref{eq:linkage1} can thus be interpreted as the greedy  maximization of modularity at maximum resolution.

\end{document}